\documentclass[prb,onecolumn,showpacs,superscriptaddress]{revtex4}
\usepackage{amsmath}
\usepackage{graphicx}% Include figure files

\def\figwidth{8.6cm}
\newcommand {\sech}{\ensuremath{\mathrm{sech}}}
\begin{document}
\title{Mean Field theory of the spin-Peierls transition}
\author{E. Orignac}
\affiliation{Laboratoire de Physique Th\'eorique de l'\'Ecole Normale
  Sup\'erieure CNRS UMR8549 24, Rue Lhomond 75231 Paris Cedex 05
  France} 
\author{R. Chitra} 
\affiliation{Laboratoire de Physique Th\'eorique des Liquides CNRS
  UMR7600 2, Place Jussieu Tour 16 75252 Paris Cedex 05 France}
\pacs{75.10.Pq 75.40.Cx 75.50.Ee}

\begin{abstract}
We revisit the problem of the spin-Peierls instability in a
one dimensional spin-$\frac12$ chain coupled to phonons.
The phonons are treated within the mean field approximation. 
We use bosonization techniques  to describe the gapped spin chain
 and then use the Thermodynamic
Bethe Ansatz to obtain \emph{quantitative} results for the
thermodynamics of the spin-Peierls system in a whole
range of temperature. This allows us to predict the behavior of the
specific heat and the magnetic susceptibility in the entire dimerized
phase. We  study the effect of small magnetic fields on the
transition. Moreover, we obtain
the parameters of the
 Landau-Ginzburg theory describing  this continuous phase transition
 near the critical point. 
\end{abstract}
\maketitle

The  spin-Peierls
instability\cite{pincus_spin-peierls,pytte_sp} is the magnetic
analogue of the 
Peierls instability of a one-dimensional metal.\cite{peierls_inst} 
 In the spin-Peierls instability, an antiferromagnetic 
spin-1/2 chain coupled to optical phonons develops a spin gap via  
 a static deformation (or dimerization)
 of the  lattice at zero temperature. 
In this dimerized phase,  the gain in magnetic energy resulting from 
 the
formation of the spin gap  outweighs the loss of
elastic energy due  to the static deformation. 
For a   system consisting of an array of spin chains coupled to two or three
dimensional phonons, the dimerized phase can persist for  
temperatures $0<T<T_{SP}$, where $T_{SP}$ is the spin-Peierls transition
temperature. For  $T>T_{SP}$,
 the chains are undistorted and  and one recovers gapless spin excitations. The phase
 transition at $T=T_{SP}$ between the dimerized and the uniform state
 is a second-order phase transition and  has
 been observed in a host of  quasi-one dimensional organic materials such
 as  TTFCuS$_4$C$_4$(CF$_3$)$_4$ and TTFAuS$_4$C$_4$(CF$_3$)$_4$
 (also known as TTFCuBDT and TTFAuBDT)\cite{jacobs_ttf_sP},
 MEM-(TCNQ)$_2$\cite{huizinga_spinpeierls},(TMTTF)$_2$PF$_6$,
 (TMTTF)$_2$AsF$_6$\cite{maaroufi_spinpeierls,pouget_tmttf2asf6,laversanne_tmttf2asf6}, and 
 (BCPTTF)$_2$PF$_6$.\cite{liu_ttf}    
 The discovery of the inorganic material\cite{hase_cugeo3}  CuGeO$_3$, spurred
further activity in this domain as this system is more convenient for
neutron scattering studies. 

 From the theory point of view, most of the treatments consider the
phonons as
static.\cite{pincus_spin-peierls,pytte_sp,cross_spinpeierls} 
This mean field treatment of the phonons is expected to work when the
phonon frequency can be neglected (adiabatic limit) compared to the
spin gap. 
Such an approach is thus better suited to the softer 
organic materials than to  
the inorganic compound CuGeO$_3$ where the frequency of the phonon
driving the transition is not small compared to the spin
gap.\cite{pouget2004}
However, a common feature of the spin-Peierls transition in all
the spin-Peierls 
compounds, is that some data  indicate a BCS type mean field behavior 
of the thermodynamic quantities near 
the transition.\cite{pouget_spinpeierls_data} 
For instance, a BCS type
relationship\cite{boucher_cugeo3_review}, $\Delta /k_B T_{SP} =
1.76$, between
the zero temperature spin gap and the spin Peierls transition temperature
$T_{SP}$ has been observed in CuGeO$_3$, and it was 
used to argue that the transition in this material could also be described
within mean field theory. However, the exact nature of the transition in
CuGeO$_3$ is still disputed.\cite{birgeneau_sP_tricritical} In
particular, no phonon softening was observed near the
transition\cite{braden_no_soft_cugeo3} in disagreement with the
mean-field
scenario.\cite{pincus_spin-peierls,pytte_sp,cross_spinpeierls} 
The absence of phonon
softening could be attributed to the  high frequency
of the phonons coupled to the spin
excitations.\cite{gros98_no_soft_rpa} Other discrepancies with the
mean-field scenario are discussed in
Ref.~\onlinecite{pouget_spinpeierls_data}. Nevertheless, 
despite these deviations,
phenomenological  Landau-Ginzburg theories can be used to some extent 
to fit the critical behavior of
CuGeO$_3$.\cite{birgeneau_sP_tricritical} Therefore, it is important 
to develop a more quantitative description of the mean field theory 
of the spin-Peierls transition, particularly in the gapped phase 
in order to have a more reliable comparison of the predictions of the
mean-field scenario with experimental data on the spin Peierls
materials.

In the first theoretical approaches to the spin-Peierls transitions\cite{pincus_spin-peierls,pytte_sp},  the spin chain was mapped
onto a model of interacting one-dimensional spinless fermions by the
Jordan-Wigner transformation\cite{jordan_transformation} and the
interactions between the fermions were either
neglected\cite{pincus_spin-peierls} or treated in the Hartree-Fock
approximation.\cite{pytte_sp}   
Later, in Ref.~\onlinecite{cross_spinpeierls}, 
though the phonons were still treated at the mean field level,
 the spin chain was described using   bosonization
 which correctly describes  the quantum critical behavior of the pure  spin
$\frac12$
chain.\cite{luther_spin1/2,haldane_xxzchain,haldane_dimerized,nijs_equivalence}
A linear response treatment of the spin-phonon coupling  resulted  
 in  a much improved estimation of dependence of the transition
temperature on the spin phonon interaction. 
Furthermore,  a Landau Ginzburg expansion was developed
 to study the vicinity of the transition.

However, in contrast to Refs.~\onlinecite{pincus_spin-peierls,pytte_sp}  no
prediction could be made  for the
thermodynamics in the dimerized phase. There are two reasons for
this. First, in Ref.~\onlinecite{cross_spinpeierls}, the dimerized phase 
is not described by a model of noninteracting fermions with a gap as
in Ref.~\onlinecite{pincus_spin-peierls}, but by a more complicated massive
sine-Gordon model.\cite{rajaraman_instanton} 
Second, in bosonization, although the expressions of 
the lattice operators in terms of  sine-Gordon fields are known, the
amplitudes in this expressions were unknown and have 
been determined quantitatively only
recently.\cite{lukyanov_ampl_xxz,affleck_ampl_xxz,lukyanov_xxz_asymptotics}  
Since the thermodynamics of the massive sine-Gordon model is now
understood\cite{fowler81_tba_sG,fowler82_Cv_sG,johnson86_sG_thermo,destri92_nostrings,destri95_tba_inteq,
destri97_tba_inteq}, and  the
exact expression of the gap in the sine-Gordon theory is known 
\cite{zamolodchikov95_gap_sinegordon}, it is  now
possible to study the thermodynamics of the dimerized spin-Peierls
phase within the mean field approximation, as well as study the zero
temperature properties. 

In this paper, we  will use the above developments to revisit the problem
of the spin-Peierls transition in the adiabatic approximation for the phonons. 
 Our  methodology and results are also applicable
to the chain mean field theory of quasi-one dimensional antiferromagnets since the latter
presents a formal analogy to the theory of the spin-Peierls
state.\cite{schulz_q1daf,wang97_chain_mf,essler97_ff_sg,irkhin00_chain_mf,bocquet02_chain_rpa,zheludev03_bacu2si2o7}  
The paper is organized as follows:
In Sec.~\ref{sec:model}, we present the model and its bosonized version. We obtain
analytical results 
results for the spin-Peierls temperature $T_{SP}$ and the total energy and gap at
 zero temperature.
In Sec.~\ref{sec:tba-meanfield},
 the Thermodynamic
Bethe Ansatz  is used to study the  thermodynamic properties of the dimerized chain at
finite temperature. We obtain  various results for the gap, the
dimerization, the specific heat  and the static magnetic susceptibility
 for  an entire range of temperatures smaller than the bandwidth.
 We use these results to  derive  the Ginzburg Landau
functional describing the mean field transition and study the behavior
of the correlation length near the transition. 

\section{Model}\label{sec:model} 
Within a mean field treatment of the phonons, 
the full Hamiltonian describing the coupling of the lattice to the spin
chain is given by
\begin{eqnarray}
  \label{eq:ham-lat}
  H=\sum_n \left[ \frac{K}{2} \langle u\rangle ^2 + J(1+(-)^n \lambda
    \langle u\rangle) {\bf
      S}_n\cdot {\bf S}_{n+1}\right],   
\end{eqnarray}
where $\langle u \rangle$ is the mean-field displacement along the chain and the ${\bf S}_i$
are spin-$\frac12$ operators. $K$ is the elastic constant for lattice
deformations and $\lambda$ is  a parameter related to the amplitude of the
spin phonon coupling.
For the sake of clarity, we
introduce the dimensionless variable   $\delta = \lambda \langle
u \rangle$, and the reduced elastic constant $\bar{K}= K/\lambda^2$. 
To obtain the physics of the spin-Peierls transition, 
 we need to evaluate the mean field displacement
$\langle u \rangle$, or equivalently, the parameter $\delta$ in a
self- consistent manner i.e., the value of $\delta$ which minimizes the
free energy at any given temperature $T$.  In the spin-Peierls phase $T \leq
T_{SP}$, $\delta(T) \neq 0$    and $\delta(T) =0$ for all temperatures $T>T_{SP}$.
A self-consistent evaluation of  $\delta (T)$ then permits a
systematic calculation of  various properties of the spin chain in a rather straightforward manner.
Since we are mainly interested in the low energy/long wavelength physics of
the
spin Peierls system, we  study the Hamiltonian (\ref{eq:ham-lat}) in
the continuum limit.
In this limit, the continuum Hamiltonian reads:  
\begin{equation}\label{eq:ham-cont}
H= \int dx \frac k 2 \delta^2  + H_s
\end{equation}
where,
 $k=\bar{K}/a$, $a$ is the lattice  spacing and $H_s$ is the continuum approximation  of the spin
 Hamiltonian. 
Using standard  
bosonization
techniques\cite{luther_spin1/2,nijs_equivalence,haldane_dimerized,schulz_houches_revue},
the continuum spin Hamiltonian is found to be :
\begin{eqnarray}
  \label{eq:ham}
  H_{s}= u \int \frac{dx}{2\pi} \left[(\pi
    \Pi)^2+(\partial_x \phi)^2\right] -\frac{2g\delta}{(2\pi
    a)^2} \int dx \cos \sqrt{2}\phi 
\end{eqnarray}
where, the fields $\phi$ and $\Pi$ are canonically conjugate to each
other (i.e. $[\phi(x),\Pi(x')]=i\delta(x-x')$). 
The  velocity of the bosonic excitations defined by the field $\phi$
is $u=\frac \pi 2 Ja$  
and  $g$ is an amplitude  proportional to the exchange interaction
$J$. The Hamiltonian (\ref{eq:ham}) is  the well known  sine-Gordon model
 with  $\beta^2=2\pi$. 
In Eq.~(\ref{eq:ham}), we have omitted a 
marginally irrelevant term.\cite{black_equ} We will
come back later to the effect of this term on the properties of the
system.  In the Hamiltonian
(\ref{eq:ham}), a non-zero dimerization $\delta$ 
induces the relevant operator,  $\cos \sqrt{2}
\phi$  of dimension $\frac12$. This results in a gap $\Delta \sim
\delta^{2/3}$, and a diminution of the  ground state energy\cite{cross_spinpeierls,haldane_dimerized} $E_s(\delta)-E_s(0) \sim
-\delta^{4/3}$. For small
$\delta$, this reduction of magnetic energy compensates the loss of
elastic energy in (\ref{eq:ham-lat}), resulting in a dimerized state
at $T=0$. 
Until recently, the proportionality  constant
 between $g$ and $J$ was unknown, thus preventing a quantitative estimation
 of the magnetic energy $E(\delta)$  and
 hence the correct value of the spin gap $\Delta$. 
Consequently, only  exponents 
could be predicted from the above  mean field description, and no
prediction could be made for the thermodynamics of the system below
the spin Peierls transition temperature. However, 
recent developments in integrable systems and bosonization, now permit
a precise determination of  
the amplitudes in the continuum
theory.\cite{lukyanov_ampl_xxz,affleck_ampl_xxz,lukyanov_xxz_asymptotics}  
A correct mapping of the lattice spin model onto its bosonized version
fixes the amplitude  $g$ in (\ref{eq:ham})
:  
\begin{eqnarray}
  \label{eq:g-def}
  g=6 J \left(\frac \pi 2\right)^{1/4} a. 
\end{eqnarray}
Although the present paper focuses on the spin-Peierls system,
we reiterate that the approach outlined below, is also applicable to the chain mean-field theory
of quasi-one dimensional
antiferromagnets\cite{schulz_q1daf,wang97_chain_mf,essler97_ff_sg,bocquet02_chain_rpa,irkhin00_chain_mf}
as long as  marginal operators are neglected. In the case of the
antiferromagnet, the magnetization $m$ and the inverse of the interchain exchange term 
$J_\perp^{-1}$  play  the role of the dimerization and the
elastic constant respectively.  
We now analyze the full Hamiltonian (\ref{eq:ham-cont}) in certain limits.
 
\subsection{Zero temperature limit}
\label{sec:zeroT}

At zero temperature, the dimerization $\delta$ is non-zero, resulting
in a gap for spin
excitations. 
As mentioned earlier, the precise mapping of the spin lattice model onto
its continuum version, the sine-Gordon model, yields exact expressions for the
gap and the total energy of the spin
system.\cite{zamolodchikov95_gap_sinegordon} In this model, the lowest energy
excitation  is a soliton\cite{rajaraman_instanton} 
and using  (\ref{eq:g-def}), its
mass is given by
\begin{eqnarray}
  \label{eq:mass}
  M&=&\frac 2 {\sqrt{\pi}} \frac{\Gamma(1/6)}{\Gamma(2/3)}
  \left[\frac{\Gamma(3/4)}{\Gamma(1/4)} \frac 3 {\pi^2} \left(\frac
      \pi 2\right)^{1/4}\delta\right]^{2/3}. 
\end{eqnarray}
Besides the  soliton and the corresponding antisoliton 
excitations, the sine-Gordon model at
$\beta_{SG}^2=2\pi$ possess two other excitations, a light breather
with a mass $M$ and a heavy breather with a mass $M\sqrt{3}$. The
soliton, the antisoliton and the light breather together form a SU(2) spin triplet
while the heavy breather forms a SU(2)
singlet.\cite{affleck86_dimerized,tsvelik92_dimerized,uhrig_dimerized}    
The gap to the lowest energy excitation or
equivalently, the singlet-triplet gap, is
\begin{equation}
  \Delta = \frac u a M \simeq 1.723 J \delta^{2/3}
\end{equation}

A comparison of this predicted value with the real gap of the spin
lattice system calculated numerically using the density matrix
renormalization group was done in Ref.~\onlinecite{orignac04_spingap} and a 
reasonably 
 good accord was found. 
The  knowledge of the soliton mass~(\ref{eq:mass}) also yields the ground
state  energy  per spin of the dimerized spin chain\cite{zamolodchikov95_gap_sinegordon}:
\begin{equation}
  \label{eq:energy-LZ}
  E_s(\delta)=-\frac \pi 2 J \frac{M^2}{4} \tan \frac \pi 6 \simeq 
  -0.2728 J \delta^{4/3},
\end{equation}
\noindent  which is in  reasonable agreement with numerics.
\cite{orignac04_spingap} 
To obtain the effective dimerization $\delta$ at zero temperature, 
we need to minimize the
total ground state energy  per unit spin of the spin lattice system,
$E= {\bar K}/2\delta^2+E_s(\delta)$ 
which leads  to the following results:
\begin{widetext}

\begin{eqnarray}
  \label{eq:zeroT}
  \Delta&=& \frac{2 \sqrt{\pi}}{3\sqrt{3}}
  \left(\frac{\Gamma(1/6)}{\Gamma(2/3)}\right)^3
  \left[\frac{\Gamma(3/4)}{\Gamma(1/4)} \frac 3 {\pi^2} \left(\frac
      \pi 2 \right)^{1/4}\right]^2 \frac{J^2}{{\bar K}}\simeq 0.627
  \frac{J^2}{\bar K} \nonumber \\
  \delta &=& \left(\frac{2}{3\sqrt{3}}\right)^{3/2} \left(\frac{\Gamma(1/6)}{\Gamma(2/3)}\right)^3
  \left[\frac{\Gamma(3/4)}{\Gamma(1/4)} \frac 3 {\pi^2} \left(\frac
      \pi 2 \right)^{1/4}\right]^2
  \left(\frac{J}{\bar K}\right)^{3/2}\simeq
  0.219\left(\frac{J}{\bar K}\right)^{3/2}  \nonumber \\
  E&=&
  -\frac{1}{\pi^7\sqrt{3}}\left(\frac{\Gamma(1/6)}{\Gamma(2/3)}\right)^6
\left(\frac{\Gamma(3/4)}{\Gamma(1/4)}\right)^4  \frac{J^3}{\bar K^2}\simeq  -0.012
\frac{J^3}{\bar K^2} 
\end{eqnarray}
\end{widetext}
We note that the ratio $J^2/{\bar K}$  can be identified with the coupling constant
$\lambda_{CF}$ used in Ref.~\onlinecite{cross_spinpeierls}, 
and  we indeed have the same exponents
 for the dependence of the gap on the coupling constant
$\lambda_{CF}$ as in Ref.~\onlinecite{cross_spinpeierls}. 
However, the prefactors in
Eqs.~(\ref{eq:zeroT}) could not be obtained in
Ref.~\onlinecite{cross_spinpeierls}.

 The continuum approximation underlying
our mean field theory is valid when the zero temperature 
 correlation length 
 is much larger than the lattice spacing i.e. when $\Delta
\ll J$. Clearly, this requires that $J\ll {\bar K}$ i.e., a sufficiently
rigid lattice. This criterion leads to 
$\delta(T=0)\ll 1$.   
We note that relations similar to the
ones in (\ref{eq:zeroT}) have been  previously obtained  in the context
of the chain mean field theory of quasi-1D
antiferromagnets.\cite{zheludev03_bacu2si2o7}    
In reality, the results of (\ref{eq:zeroT}) 
are slightly modified by the presence of a
marginally irrelevant term in the continuum Hamiltonian for the spin
system.\cite{kadanoff_ashkin,black_equ,kohmoto81_ashkin} When the
marginal interaction is taken into account, it is found that in the
spin-Peierls case $\Delta \sim \delta^{2/3}|\ln \delta|^{-1/2}$,
whereas in the case of the antiferromagnet $\Delta \sim \delta^{2/3}
|\ln \delta|^{1/6}$ i.e. the marginally irrelevant term frustrates the
dimerization and favors antiferromagnetic ordering. 
The marginal interaction is eliminated in the
$J_1$--$J_2$ chain at its critical point\cite{affleck_log_corr,chitra95_dmrg_frustrated} $J_2/J_1\simeq
0.24$.  With an additional
dimerization of the nearest neighbor exchange\cite{chitra95_dmrg_frustrated} in this critical chain, the gap  $\Delta=1.76
\delta^{2/3}$.  Moreover, in the
absence of dimerization\cite{okamoto_frustrated}, the spin velocity at the critical point 
 is found to be $1.1936 J_1
a$. 
Generalizing the results obtained above, we find that the following amplitude
$g=0.806 \pi^2 J_1 a$  should  be used in the bosonized Hamiltonian
(\ref{eq:ham})  in order to describe the $J_1- J_2$ chain with
$J_2/J_1= 0.2411$.
The resulting energy gain from dimerization is then
$E_s/J_1=-0.3745\delta^{4/3}$  and  one obtains
the following zero temperature results:
\begin{eqnarray}
  \label{eq:J1-J2case}
\Delta & =& 0.879\frac{J_1^2}{{\bar K}}\nonumber \\  
\delta & =& 0.353\left(\frac{J_1}{\bar K}\right)^{3/2} \nonumber \\
E& =&-0.093  \frac{J_1^3}{{\bar K}^2}
\end{eqnarray}
Comparing (\ref{eq:J1-J2case}) and (\ref{eq:zeroT}), we see that the
introduction of $J_2$ 
results in a strong enhancement of the zero temperature gap and of the
dimerization, in agreement with a scenario proposed in
Ref.~\onlinecite{castilla_cugeo3} for the spin-Peierls transition in
CuGeO$_3$.

\subsection{Transition temperature}
\label{sec:Tsp-perturb}

In this section, we redo the calculation of
Ref.~\onlinecite{cross_spinpeierls} yielding the
spin-Peierls transition temperature.
For  any temperature, a 
self-consistent treatment of the problem requires a calculation of
the free energy as a function of the dimerization $\delta$ taken as a
variational parameter, followed by a  minimization  with respect to $\delta$.
Unlike the zero temperature case, calculating
the  free energy for arbitrary temperatures
 requires the use of Thermodynamic Bethe Ansatz
 techniques\cite{fowler81_tba_sG,fowler82_Cv_sG,johnson86_sG_thermo}, and no
 closed analytic expressions can be obtained. 
However, to calculate the spin Peierls transition
temperature, this full treatment is not
required.\cite{cross_spinpeierls}  
Indeed, close to the spin-Peierls transition, the order parameter
$\delta$ becomes small and a second order  
perturbation theory in $\delta$, is sufficient to evaluate the leading
behavior of the variational free energy.  
A straightforward  perturbative development in $\delta$ of the
Matsubara imaginary time path integral   gives the following
expression for the
free energy of the sine-Gordon model:
\begin{widetext}
\begin{eqnarray}
  \label{eq:high-T-exp}
  F=-\frac{\pi}{6 u} T^2 -\frac 1 4 \frac{\pi
    a}{\beta u} \left(\frac{2g\delta}{(2\pi
      a)^2}\right)^2 \int_{-\infty}^\infty dx \int_0^\beta d\tau
  \frac{\sqrt{2}}{\sqrt{\cosh \frac{2\pi x}{\beta u}-\cos \frac {2\pi
        \tau}{\beta}}} +o(\delta^2)
\end{eqnarray}
\end{widetext}
Note that the first term in this expression is just the free energy
of a non-interacting  Bose gas in one dimension.   
 Using Eq.~(8.12.4) in 
 Ref.~\onlinecite{abramowitz_math_functions}  to integrate over the
space variable $x$, we obtain
\begin{eqnarray}
  F_s(T,\delta)=-\frac{\pi}{6 u} T^2 -\frac a 4  \left(\frac{2g\delta}{(2\pi
      a)^2}\right)^2 \int_0^\beta \pi P_{-1/2}\left(-\cos \frac {2\pi
        \tau}{\beta}\right) d\tau,  
\end{eqnarray}
where the function $P_{-1/2}$ is  a Legendre function. A final
integration over $\tau$ using Eq.~(8.14.16) in
Ref.~\onlinecite{abramowitz_math_functions} leads to: 
\begin{eqnarray}
  \label{eq:hTexp-final}
  F_s(T,\delta)&=&-\frac{\pi}{6 u} T^2-\frac{a}{4}\left(\frac {2 g \delta}{(2
      \pi
      a)^2}\right)^2\frac{\pi^2}{\Gamma(3/4)^4 T} \nonumber \\
   &=& -\frac{\pi}{6 u} T^2-\frac{9 J^2 \delta^2}{4 \pi^2 \Gamma(\frac34)^4
     a T}\left(\frac \pi 2 \right)^{1/2}
\end{eqnarray}
The full mean field variational free energy is
$F_{MF}(T,\delta)= \frac k 2 \delta^2 + F_s(T,\delta)=C\delta^2
+o(\delta^2)$. When $C>0$, which is obviously the case for high
temperature, the mean field free energy has a minimum for
$\delta=0$ and for $C<0$,  the energy is minimized by a state with non-zero
dimerization. Therefore,  the spin-Peierls transition
temperature is  fixed by the condition $C=0$
and  using (\ref{eq:hTexp-final}), we obtain
\begin{eqnarray}
\label{eq:T-SP}
  T_{SP} &=&\frac{9}{2\pi^2 \Gamma(3/4)^4}
  \left(\frac \pi 2 \right)^{1/2} \frac{J^2}{\bar K} = 0.25342 \frac{J^2}{\bar K}
\end{eqnarray}
Note that the  validity of the continuum description  requires that
$T_{SP}\ll J$. 
In Ref.~\onlinecite{cross_spinpeierls}, the same dependence of
$T_{SP}$ on
$J^2/{\bar K}$  (up to the
prefactor) was derived using an equivalent response function
formalism. Comparing the two expressions, we observe that in
Ref.\cite{cross_spinpeierls}, the  transition
temperature $T_{SP}\simeq 1.01 J^2/{\bar K}$ obtained there is a gross overestimate of
$T_{SP}$ highlighting the importance of having correct amplitudes in the
bosonized theory. 
Comparing Eqs.~(\ref{eq:zeroT}) and (\ref{eq:T-SP}), we note  that the
ratio: 
\begin{equation}\label{eq:bcs-ratio}
\frac{\Delta(T=0)}{T_{SP}} \simeq 2.47
\end{equation}
 is independent of the various
coupling constants present in the theory. This ratio is in accord with
values obtained by numerical studies of the spin-Peierls problem.\cite{pouget_spinpeierls_data}  The existence of such an 
universal ratio is reminiscent
 of  the BCS  mean field theory for superconductivity\cite{tinkham_book_superconductors} where  the ratio
of the superconducting gap and transition temperature is approximately $1.76$
. In fact, one can use the Jordan-Wigner
transformation \cite{jordan_transformation} to map the Heisenberg spin
chain onto a chain of interacting spinless fermions. Neglecting the
interactions\cite{pincus_spin-peierls}, the resulting theory presents
a formal similarity with the BCS theory\cite{tinkham_book_superconductors}, which leads
one to anticipate an universal  ratio.  We note, however, that the fact that
the spinless fermions  theory is strongly
interacting  renormalizes the BCS ratio away from the non-interacting
value $1.76$. In particular, as already discussed in
Ref.~\onlinecite{pouget_spinpeierls_data}, the observation of a ratio of
$1.76$ between the zero temperature gap and the transition temperature
in $\mathrm{CuGeO_3}$ \emph{cannot} be taken as an indication of
adiabatic behavior in this compound.   
As discussed earlier,  the results of
(\ref{eq:zeroT}) were obtained neglecting the logarithmic corrections
induced by a marginally irrelevant interaction. These marginal
interactions affect
 the dependence of the
gap and the ground state energy on the dimerization, particularly for
$\delta \ll 1$ and at  finite temperatures induce 
 logarithmic
corrections in response functions.\cite{starykh97_logs} This inhibits
a precise estimation of the BCS like ratio especially in systems
 with  a small
dimerization at low temperatures.   

 For the
next nearest  neighbor chain  
with a critical coupling  $J_{2c} = 0.2411$, where 
logarithmic corrections vanish,   $\Delta= 1.5386 J_1^2/{\bar K}$ and
$T_{SP}= 0.623 J_1^2/{\bar K} $.
Note that these values respect the BCS relation~(\ref{eq:bcs-ratio}).
In the light of the preceding discussion, it is interesting to
note that a small change in the velocity and the coefficient
of the
sine-Gordon term, leads to a big change in the gap and the
spin-Peierls
temperature. This implies that the  frustration engendered by $J_2$
enhances fluctuations towards spontaneous dimerization, hence
favoring
 the formation of the spin-Peierls state.

\subsection{Effect of logarithmic corrections}\label{sec:effect-log-corr}
We now discuss the effect of the marginally
irrelevant operator $\cos \sqrt{8}\phi$, neglected in the preceding
sections. This operator
is known to induce 
logarithmic corrections in the dimerization
gap\cite{black_equ,kadanoff_ashkin,affleck_log_corr} as well as
in  response
functions.\cite{affleck_log_corr,giamarchi_logs,barzykin99_logs,barzykin_log_temperature}
This results in a modification of Eqs.~(\ref{eq:zeroT}) and
(\ref{eq:T-SP}) and consequently deviations from the BCS like  ratio
(\ref{eq:bcs-ratio}). Including these corrections, the gap at $T=0$ and the
ground state energy  are:
\begin{eqnarray}
  \label{eq:gap-logs}
  \Delta &\sim& J \frac{\delta^{2/3}}{|\ln \delta|^{1/2}} \\
  \label{eq:ground-log}
  E_0 &\sim& -J \frac{\delta^{4/3}}{|\ln \delta|} +\frac{\bar K} 2 \delta^2
\end{eqnarray}
Minimizing the ground state energy with respect to $\delta$, one
finds:
\begin{eqnarray}
  \label{eq:dimer-log}
  \delta^{2/3}|\ln \delta| \sim \frac J {\bar K} 
\end{eqnarray}
For the transition temperature, it can be shown following
Ref.~\onlinecite{barzykin_log_temperature} that the susceptibility
associated with the dimerization operator is corrected by a
logarithmic factor so that:
\begin{eqnarray}
  \label{eq:resp-log}
  \chi_{d}(T)\sim \frac 1 T \left(\ln \frac J T\right)^{-3/2} 
\end{eqnarray}
With this result, the equation (\ref{eq:T-SP}) is modified into:
\begin{eqnarray}
  \label{eq:Tsp-log}
  T_{SP} \left(\ln \frac J {T_{SP}}\right)^{3/2} \sim \frac{J^2}{\bar K}  
\end{eqnarray}

We see from Eqs.~(\ref{eq:dimer-log}) and (\ref{eq:Tsp-log}) that the
effect of logarithmic corrections is to decrease  the spin-Peierls
transition temperature and the zero temperature gap. In contrast, in
the case of the N\'eel state, these logarithmic corrections enhance
the transition temperature and the order parameter.\cite{schulz_q1daf,essler97_ff_sg,zheludev03_bacu2si2o7,irkhin00_chain_mf,bocquet02_chain_rpa}

The equation~(\ref{eq:dimer-log}) leads to:
\begin{eqnarray}
  \label{eq:delta-logcor}
  \delta \sim \left(\frac{\frac{J}{\bar K}}{\frac 3 2 \left|\ln \frac{J}{\bar K}\right|}\right)^{3/2},
\end{eqnarray}
\noindent resulting in a gap:
\begin{eqnarray}
  \label{eq:gap-logcor}
  \Delta \sim \frac{J^2}{\bar K}\frac 1 {\left|\ln \frac{J}{\bar K}\right|^{3/2}},
\end{eqnarray} 
Solving for $T_{SP}$ in Eq.~(\ref{eq:Tsp-log}), and comparing with
Eq.~(\ref{eq:gap-logcor}), we find that to lowest order a BCS type
relation holds. This relation however is obtained by retaining only
the lowest order logarithmic corrections.  
\section{Thermodynamic Bethe Ansatz Mean Field theory}
\label{sec:tba-meanfield}

\subsection{Mean Field equations}
\label{sec:mf-equations}

We now focus on  the mean field theory for the spin lattice system at
arbitrary temperatures.
We use the  Thermodynamic Bethe Ansatz (TBA) 
as a tool to  evaluate the
finite temperature free energy of the sine-Gordon
Hamiltonian~(\ref{eq:ham}). The  TBA treatment of the generic
sine-Gordon model with a relevant term $\cos \beta \phi$ has been formulated
using the string hypothesis in 
Refs.~\onlinecite{fowler81_tba_sG,fowler82_Cv_sG,johnson86_sG_thermo}. 
In general, this leads to an infinite number
of coupled integral equations for the various pseudoenergies. However,
at the so called  reflectionless points\cite{dorey_smatrix_review},
defined by 
$\beta^2=\frac{8\pi}{n}$, where $n$ is an integer, 
the number of independent  integral equations
becomes finite.\cite{johnson86_sG_thermo} Numerical methods can then
be used to solve these integral equations and deduce the thermodynamics.
The case of the dimerized spin 
chain with $\beta^2=2\pi$ 
falls into this category and the TBA equations of
Refs.~\onlinecite{fowler81_tba_sG,fowler82_Cv_sG,johnson86_sG_thermo} 
 can be used to  calculate the free energy. For generic values of $\beta$ 
 away from the reflectionless points,  the  general formalism developed
 by Destri and de
 Vega\cite{destri92_nostrings,destri95_tba_inteq,destri97_tba_inteq}
 is more appropriate than the string approach. This latter method has
been used  successfully to obtain  the thermodynamic properties of copper
 benzoate\cite{essler99_cubenzoate} in a magnetic field. This
 approach  can be used to study the thermodynamics of
 the generalized spin-Peierls
 transition\cite{nagaosa_lattice_ladder,calemczuk_heat_ladder} 
 or the antiferromagnetic transition\cite{giamarchi_coupled_ladders}
 in systems of coupled spin ladders in a  magnetic field.  
Before we embark on an application of the TBA method to the spin
Peierls system, we note  that in
Refs.~\onlinecite{fowler81_tba_sG,fowler82_Cv_sG,johnson86_sG_thermo} 
the free energy is
taken to be zero at zero temperature. However, since  our reference state is
the undimerized chain,  we must add 
the zero temperature
dimerization energy to the free energy calculated using  the TBA of
Refs.~\onlinecite{fowler81_tba_sG,fowler82_Cv_sG,johnson86_sG_thermo}.   
Using the approach outlined in
Refs.~\onlinecite{fowler81_tba_sG,fowler82_Cv_sG,johnson86_sG_thermo}, 
we find that in our case, the sine-Gordon free
energy reads:
\begin{widetext}
\begin{equation}
  \label{eq:TBA-equations}
  F_s(T,\delta) =-\frac{T}{2\pi u} \int^\infty_{-\infty}
 d\theta \Delta \cosh \theta \left[ 3 \ln
  (1+e^{-\epsilon_1(\theta)/T}) +\sqrt{3} \ln (1
  +e^{-\epsilon_2(\theta)/T}) \right] -\frac{u}{a^2} \tan \frac
\pi 6 \frac{M^2}{4}\end{equation}
\end{widetext}
\noindent
where  the pseudoenergies 
$\epsilon_1(\theta)$ and $\epsilon_2(\theta)$ are
self-consistently determined by the following integral  equations:
\begin{widetext}
\begin{eqnarray}
  \epsilon_1(\theta)&=&\Delta \cosh \theta + \frac {3T}{2\pi}
  \int_{-\infty}^{\infty} d\theta' K_{11}(\theta-\theta') \ln
  (1+e^{-\epsilon_1(\theta')/T})  + \frac
  {T}{2\pi}\int_{-\infty}^{\infty} d\theta' K_{12}(\theta-\theta') \ln
  (1+e^{-\epsilon_2(\theta')/T}), \nonumber \\ 
\epsilon_2(\theta)&=&\Delta \sqrt{3} \cosh \theta + \frac {3T}{2\pi}
  \int_{-\infty}^{\infty} d\theta' K_{12}(\theta-\theta') \ln
  (1+e^{-\epsilon_1(\theta')/T}) + \frac
  {T}{2\pi}\int_{-\infty}^{\infty} d\theta' K_{22}(\theta-\theta') \ln
  (1+e^{-\epsilon_2(\theta')/T}),  
\end{eqnarray}
\end{widetext}
The integral kernels are given by:
\begin{eqnarray}
  \label{eq:kernels}
  K_{11}(\theta)&=&\frac{2\sin  \frac \pi 3 \cosh \theta}{\sinh^2\theta+\sin^2\frac
    \pi 3} \nonumber \\
  K_{12}(\theta)&=&\frac{2\sin  \frac \pi 6 \cosh
    \theta}{\sinh^2\theta+\sin^2\frac \pi 6} + \frac{2 \cosh
    \theta\sin  \frac \pi 2}{\sin^2 \frac \pi 2+\sinh^2 \theta}\nonumber  \\ 
  K_{22}(\theta)&=&3K_{11}(\theta)
\end{eqnarray}
The pseudoenergies $\epsilon_{1,2}$ have a transparent physical
interpretation\cite{saleur_houches}: 
$\epsilon_1(\theta)$ represents the dressed energy of the solitons and
of the light breather (which have identical masses at the
$\beta^2=2\pi$ point), whereas the pseudoenergy $\epsilon_2(\theta)$
represents the dressed energy of the heavy breather. In fact, because
scattering is diagonal, 
Eqs.~(\ref{eq:TBA-equations})--(\ref{eq:kernels}) can be easily
re-derived using the approach outlined  in Ref.~\onlinecite{saleur_houches}.    
It is useful to recast the dimensionless free energy $f=a F/J$
 in terms of  the scaled energies
 $\bar{\epsilon}_i=\epsilon_i/J$ and the
reduced temperature $\bar{T}=T/J$. 
\begin{widetext}
\begin{eqnarray}
\label{reduced-f}
  f&=&\frac{\bar{K}}{2J} \delta^2 -\frac{\bar{T}}{2\pi}\int d\theta  M \cosh \theta \left[ 3 \ln
  (1+e^{-\bar{\epsilon_1}(\theta)/\bar{T}})  +\sqrt{3} \ln (1
  +e^{-\bar{\epsilon_2}(\theta)/\bar{T}}) \right]
-\frac{\pi}{8\sqrt{3}} M^2 \\
\label{inteq1}
\bar{\epsilon}_1(\theta)&=&\frac{\pi}{2} M \cosh \theta +
\frac{3\bar{T}}{2\pi} \int_{-\infty}^{\infty} d\theta' K_{11}(\theta-\theta') \ln
  (1+e^{-\bar{\epsilon_1}(\theta')/\bar{T}})  + \frac {\bar{T}}{2\pi}\int_{-\infty}^{\infty} d\theta' K_{12}(\theta-\theta') \ln
  (1+e^{-\bar{\epsilon}_2(\theta')/\bar{T}}) \\
\label{inteq2} 
\bar{\epsilon}_2(\theta)&=&\frac{\pi} 2  M \sqrt{3} \cosh \theta + \frac {3\bar{T}}{2\pi}
  \int_{-\infty}^{\infty} d\theta' K_{12}(\theta-\theta') \ln
  (1+e^{-\bar{\epsilon}_1(\theta')/\bar{T}})  + \frac {\bar{T}}{2\pi}\int_{-\infty}^{\infty} d\theta' K_{22}(\theta-\theta') \ln
  (1+e^{-\bar{\epsilon}_2(\theta')/\bar{T}})   
\end{eqnarray}
\end{widetext}
To obtain the free energy for arbitrary temperatures we need to solve
for a given  soliton mass $M$, 
the
pair of self consistent equations~(\ref{eq:TBA-equations}) 
for the dispersions $\epsilon_1$ and $\epsilon_2$.
Here, we use numerical methods~\cite{fowler82_Cv_sG} to obtain the
solutions
for any arbitrary temperature. 
We use a simple iterative procedure to solve
(\ref{inteq1})--~(\ref{inteq2}) numerically for various values of
the dimerization $\delta$ at a fixed  temperature $T$.
This  provides us with the variational free energy
energy for an entire range of  $\delta$ at fixed temperature. We then
identify the  value of $\delta$  for which the free energy is a
minimum. Such a procedure is repeated for various values of the
temperature $T$ thereby permitting us to obtain $\delta(T)$.
In particular, 
we find that the numerical estimate of $T_{SP}$ coincides perfectly
with the prediction of perturbation theory Eq.~(\ref{eq:T-SP}). This
provides a first check of the validity of our TBA mean field solution. 
In Fig.~\ref{fig:dimerization},
we plot our results for the mean field dimerization $\delta(T)$ as a 
function of the reduced temperature. 
These results for $\delta(T)$ are then used to obtain thermodynamic quantities.   
Figs.~\ref{fig:gap} and ~\ref{fig:specheat}  show a  plot the gap  and
specific heat as as function of the temperature.
In the vicinity of $T_{SP}$, the gap vanishes as   $\Delta \propto (T_{SP} -T)^{1/3}$
 and the  specific heat jumps at $T_{SP}$ as expected for a second
order transition.
For  low temperatures $T \ll  T_{SP}$ the specific heat is
exponentially suppressed by the spin gap and for   $T \geq T_{SP}$, the
specific heat is simply that
of 
the pure  Heisenberg chain:
$C_v=\frac{\pi}{3u} T=\frac{2T}{3J}$ which is the same as that for
a gas of free bosons.

\subsection{Law of corresponding states}\label{sec:law-corr-stat}
 
\begin{figure}[htbp]
  \begin{center}
    \includegraphics[width=\figwidth]{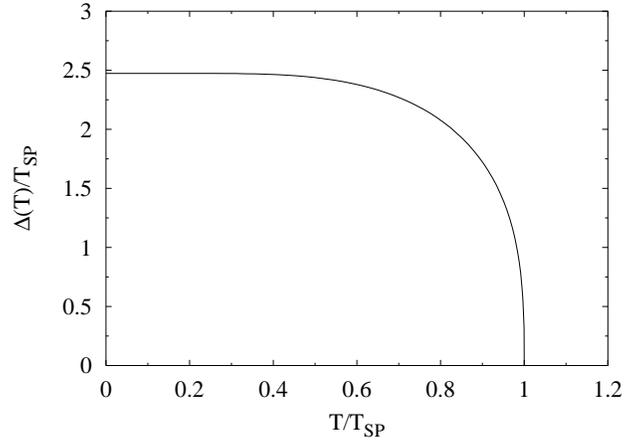}
    \caption{The dimensionless scaling function $\bar{\Delta}$ 
describing the law of
      corresponding states followed by the spin-Peierls gap. The
      universal ratio 2.47 is reached for $T<0.4T_{SP}$. For $T\to
      T_{SP}$ the scaling function $\bar{\Delta} \sim
      (1-T/T_{SP})^{1/3}$.}
    \label{fig:gap}
  \end{center}
\end{figure}

\begin{figure}[htbp]
  \begin{center}
    \includegraphics[width=\figwidth]{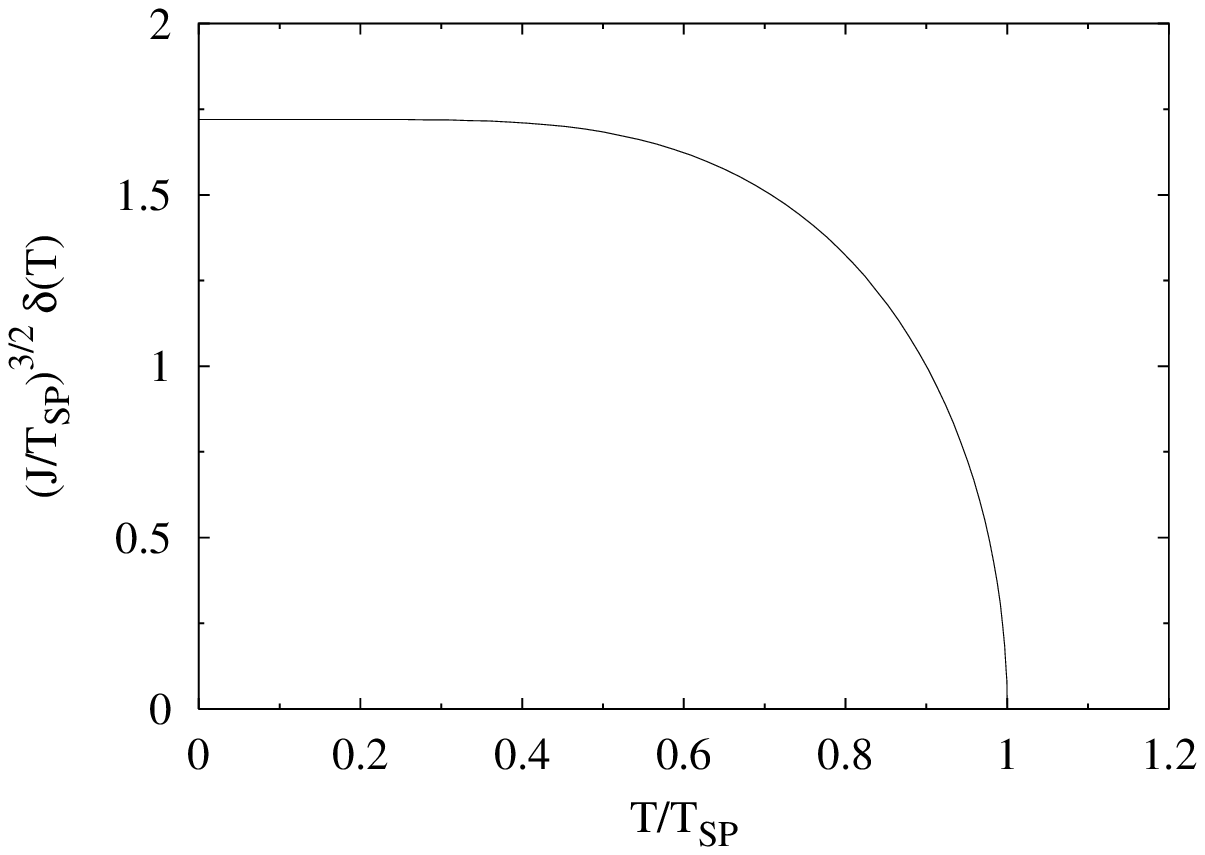}
    \caption{The dimensionless scaling function $\bar{\delta}$ 
describing the law of
      corresponding states followed by the spin-Peierls
      dimerization. The zero temperature value is reached for $T<0.4
      T_{SP}$. For $T\to
      T_{SP}$ the scaling function $\bar{\delta} \sim
      (1-T/T_{SP})^{1/2}$.}
    \label{fig:dimerization}
  \end{center}
\end{figure}

\begin{figure}[htbp]
  \begin{center}
    \includegraphics[width=\figwidth]{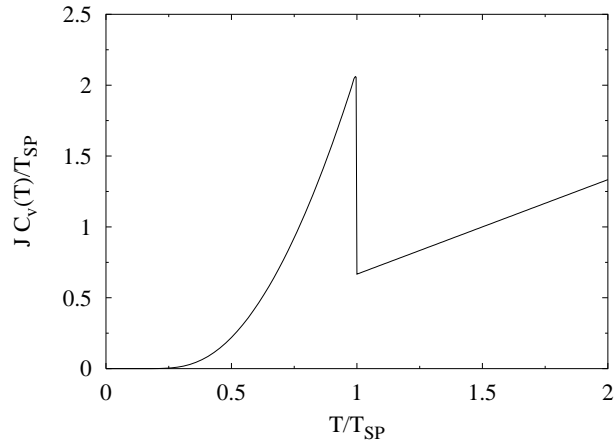} 
    \caption{The specific heat of the spin-Peierls problem in the mean
      field approximation.}
    \label{fig:specheat}
  \end{center}
\end{figure}

The mean-field
theory leads to a law of corresponding states\cite{landau_statmech} or
equivalently, scaling forms for the free energy and associated quantities.
We use the  $T=0$ result $\delta \sim (J/{\bar K})^{3/2}$ (cf. Eq.~(\ref{eq:zeroT})) 
  to rewrite the finite temperature dimerization 
 $\delta(T)=(J/{\bar K})^{3/2} \bar{\delta}(T)$. Inserting this  in
 Eq.~(\ref{eq:mass}), and in Eqs.~(\ref{inteq1})--~(\ref{inteq2}), and
 using (\ref{eq:T-SP}), it is straightforward to see that
 the pseudoenergies satisfy the scaling form  $\epsilon_i(\theta)=T_{SP}\bar{\epsilon}(T/T_{SP},\theta)$. Consequently, this implies that
  the total free energy, gap and dimerization  can be re-expressed as: 
\begin{eqnarray}
  \label{eq:corresp-states}
  F(T) &= &-\frac{T_{SP}^2}{J} f\left(\frac{T}{T_{SP}}\right)\\ 
  \delta(T)&= &\left(\frac{T_{SP}}{J}\right)^{3/2} \bar{\delta}\left(\frac{T}{T_{SP}}\right) \\
  \Delta(T)&= &T_{SP} \bar{\Delta}\left(\frac{T}{T_{SP}}\right)
\end{eqnarray}
where the functions $\bar{\Delta},\bar{\delta},f$ are universal
functions of the scaled temperature.  From Figs.~\ref{fig:gap} and
~\ref{fig:dimerization}, we see that the numerical solutions for $\bar{\Delta},\bar{\delta}$   do obey
the above scaling form.
 From the expression for  the free energy (\ref{eq:corresp-states}),
one easily
obtains the  following result  for the specific heat:
\begin{eqnarray}
  \label{eq:specif-heat}
  C_v=\frac{T_{SP}}{J} \frac{T}{T_{SP}}
  f^{\prime\prime}\left(\frac{T}{T_{SP}}\right) 
\end{eqnarray}
  As expected in a mean-field
theory, there is a jump in
the specific heat at the transition  whose magnitude is given by
\begin{eqnarray}
  \label{eq:spec-jump}
 \frac{J \Delta C_v}{T_{SP}}=f^{\prime\prime}(1^{-})-\frac 2 3=\gamma_{SP}
\end{eqnarray}
The numerical solution of the mean field equation yields
$\gamma_{SP}=1.39$. Our result for the specific heat  is shown in
Fig.~\ref{fig:specheat}. A universal 
ratio of the specific heat jump to the specific heat above the critical
temperature\cite{tinkham_book_superconductors} exists in the BCS theory, where
${\Delta C_v}/C_v(T_{SP}^+)=1.43$. Here again, the value of the ratio
$\Delta C_v/C_v(T_{SP}^{+})=2.1$ in the spin Peierls problem is
different from the BCS ratio of $1.43$ due to the strong interactions
between the Jordan-Wigner pseudofermions. We note that in experiments
on $\mathrm{CuGeO_3}$, this ratio was found to be $1.5$ or $1.6$,
which is close to the BCS
prediction.\cite{lasjaunias_heat_cugeo3} However, as in the case of
the ratio of the gap to the transition temperature, this apparent
agreement with the mean field description proves to be spurious as we
have seen that the
ratio should be near $2.1$ in a spin isotropic material.  

\subsection{Landau-Ginzburg expansion}\label{sec:LG-expans}
In this section, we use the results of the preceding sections to
obtain a simple Ginzburg Landau functional describing the vicinity of
the spin-Peierls transition.
In Ref.~\onlinecite{cross_spinpeierls}, it was shown that a soft 
Ising or $\phi^4$ theory was enough to describe the vicinity of the
transition. However, the  coefficients, in particular, that of
 the quartic term could not be entirely
calculated.
 Our
formalism permits us to obtain the leading terms in the functional
with the correct prefactors. This will be useful for more
sophisticated treatments of the transition taking into account the
fluctuations of the lattice\cite{lebellac_qft} or the role of
solitons in the thermodynamics.\cite{scalapino_q1d}  
A Landau Ginzburg expansion\cite{landau_statmech} of the variational free
energy per unit length~(\ref{eq:TBA-equations}) in the vicinity of $T_{SP}$ gives
\begin{eqnarray}
  \label{eq:LG-exp}
  F(T,\delta)=\frac{p}{2} (T-T_{SP}) \delta^2 + \frac{q}{4} \delta^4
\end{eqnarray}
The law of corresponding states
(\ref{eq:corresp-states}) leads to some constraints on the form of the
expansion.
Minimizing with respect to $\delta$ yields:
\begin{eqnarray}
  \label{eq:MF-critical}
  \delta^2(T)=\frac{p}{q}(T_{SP}-T) \nonumber \\
  F(T)=-\frac{p^2}{4q}(T-T_{SP})^2
\end{eqnarray}
The law of corresponding states (\ref{eq:corresp-states}) implies that
$p/q\sim T_{SP}^2/J^3$ and that $p^2/q \sim 1/J$. Thus, we have
$p=c_1  J^2/T_{SP}^2$ and $q=c_2 J^5/T_{SP}^4$, where
$c_1, c_2$ are dimensionless numbers.  These predictions are in
agreement with the ones obtained from the RG treatment  in
Ref.~\onlinecite{bourbonnais96_spinpeierls}. 
The dependence of $p$  can also be verified by  the
perturbation theory of Sec.~\ref{sec:Tsp-perturb}. Although the
precise value of $q$ was not obtained in Ref.~\onlinecite{cross_spinpeierls}, it
was shown  using perturbation theory
that the Landau-Ginzburg free energy had an expansion in
powers of $(J/T_{SP})^{1/2} \Delta_0/T_{SP}$, where
$\Delta_0=J\delta$. Reporting  this expansion in Eq. (5.5) of
Ref.~\onlinecite{cross_spinpeierls},  leads precisely to the
dependence of $q$ on $J$ and $T_{SP}$. Thus, the perturbative
expansion of the free energy is fully consistent with the law of
corresponding states.   
In terms of the dimensionless constants, 
\begin{eqnarray}
\delta(T)^2 & = &\frac{c_1}{c_2} \left(\frac{T_{SP}}{J}\right)^3
\left(1-\frac {T}{T_{SP}}\right)
\nonumber \\
F(T)& =& -\frac{c_1^2}{4c_2} \frac{(T-T_{SP})^2}{J} 
\end{eqnarray}
This also implies from
Eq.~(\ref{eq:mass}) that
the spin-Peierls gap vanishes as $\sim
(1-T/T_{SP})^{1/3}$ near the transition.   
This behavior is entirely consistent with our numerical results Eq.~(\ref{eq:MF-critical}).
It now suffices to calculate the constants $c_1$ and $c_2$.
A comparison with
Eq.~(\ref{eq:hTexp-final}), fixes  $c_1= 0.2534$ 
and the value of $c_2=0.02276 $ is obtained by fitting the TBA mean
field theory results for $\delta(T)$  to (\ref{eq:MF-critical}) in the
range $0.9T_{SP}<T<T_{SP}$. 
Hence  
\begin{eqnarray}
  \label{eq:coeff-quadra}
  p\simeq 0.2534 \frac{J^2}{T_{SP}^2} \nonumber \\
  q\simeq 0.0228\frac{J^5}{T_{SP}^4}
\end{eqnarray}
As a check of the correctness of the results of the Ginzburg Landau
expansion, we can compare the prediction for the  specific
heat jump from Eq.~(\ref{eq:MF-critical}), $\Delta C_v=\frac{c_1^2}{2c_2} \frac{T}{J}\simeq 1.4(1) \frac
T J$ with the value given by the TBA mean field theory in
Eq. (\ref{eq:spec-jump}) $\Delta
C_v\simeq 1.39 \frac
T J$.  The $1\%$  agreement between these two values provides a confirmation
of the correctness of our Ginzburg-Landau expansion. 
 From the behavior of the gap, we can obtain the behavior of the
magnetic correlation length $\xi_{mag}(T)$. If the gap is $\Delta$,
the magnetic correlation length at $T=0$ is $u/\Delta$. Thus,
neglecting thermal effects, we would obtain $\xi_{mag}(T)\sim J/T_{SP}
(1-T/T_{SP})^{-1/3}$. Near the transition, this correlation length
becomes much larger than the thermal correlation length $\xi_{th}=u/(2\pi T)$. This means that the exponential decay of the magnetic
correlation that would result from the gap $\Delta(T)$ is completely
masked by the thermal fluctuations which lead to a much shorter
correlation length.  
The above results are valid for the case of a uniform
dimerization $\delta$. 
In reality, near the transition, the dimerization can vary with the
spatial location.  To take into account the energy cost of these
fluctuations, a  $(\nabla \delta)^2$ term must be included 
in the Landau-Ginzburg effective
theory.
The bosonization approach allows to calculate the
coefficient
of this gradient term as outlined in App.~\ref{sec:calculation-rigidity}. 
The full Landau Ginzburg free energy is now given by
\begin{equation}\label{eq:lg}
F_L=   \int dx \left[\frac{c_0}{2} (\nabla \delta)^2 + \frac p 2
  (T-T_{SP})
  \delta^2(x) + \frac q 4  \delta^4(x)\right]
\end{equation}
where the constant $c_0$ measures the rigidity of the order parameter
$\delta$ and is given (see App.~\ref{sec:calculation-rigidity}) by: 
\begin{equation}
c_0= \frac{9}{8\pi^2} \left(\frac \pi 2\right)^{1/2}
  \frac {\beta(2)} {\Gamma(3/4)^4} \frac{J^4 a}{T^3}
\end{equation}
where $\beta(2)\simeq 0.91596\ldots$ is Catalan's
constant\cite{abramowitz_math_functions} and $p,q$ are given by (\ref{eq:coeff-quadra}).
It is interesting to note that the coefficient of the gradient term
falls as $T^3$. We note that the Ginzburg-Landau coefficients calculated in
Ref.~\onlinecite{bourbonnais96_spinpeierls} using a fermionic 
renormalization group treatment  have the same dependence on
$T_{SP}$ as in the present mean-field calculation. However, we do not
expect an agreement of the numerical prefactors as the model studied
in Ref.~\onlinecite{bourbonnais96_spinpeierls} is different from the one studied
here. 
The  structural correlation length close to the transition
 can be evaluated from (\ref{eq:lg})
and is found be to be
\begin{eqnarray}
  \label{eq:corr-length-struct}
  \xi^2(T) &=& \frac{c_0}{p(T-T_{SP})}=\left(\frac{J a}{2\pi T_{SP}}\right)^2
  \beta(2) \frac{T_{SP}}{T-T_{SP}} \nonumber \\
       &=& \xi_{th}^2 \left(\frac 2
    {\pi}\right)^2 \beta(2) \frac{T_{SP}}{T-T_{SP}}. 
\end{eqnarray}
Near the transition, $\xi\gg \xi_{th}$. This
justifies  the Landau-Ginzburg approach where  the magnetic
fluctuations  are integrated out and only  structural fluctuations 
close to the transition are retained.  As was done in
Ref.~\onlinecite{bourbonnais96_spinpeierls}, the contributions of these
structural fluctuations to the specific heat can be analyzed by the
techniques of Refs.\onlinecite{scalapino_1d_2d,scalapino_q1d}.  

\subsection{Magnetic susceptibility}\label{sec:magn-susc}

Here we consider the effect of a magnetic field on the spin-Peierls
system.
The field can close the spin triplet gap and induce incommensuration.
Here, we restrict ourselves to fields much smaller than the gap and
study their effect on the spin Peierls transition temperature and
the susceptibility.
 Using the perturbative approach \cite{cross_field_sP}
 generalizing the one of
Sec.~\ref{sec:Tsp-perturb} we find that for small magnetic fields
there is a reduction of the  transition 
 temperature $T_{SP}$ i.e., $T_{SP}(h)=T_{SP}(0)- \lambda h^2$, with $\lambda>0$. 
Using bosonization (see App.~\ref{sec:calculation-rigidity}), we find: 
\begin{eqnarray}
  \label{eq:lambda-Tsp}
  \lambda=\frac{\beta(2)}{\pi^2 T_{SP}}\simeq
  \frac{14.7}{16\pi^2 T_{SP}},
\end{eqnarray}
\noindent    This result is
in
accord with that of Ref.~\onlinecite{cross_field_sP} 
where it was found that $\lambda\simeq 14.4/(16\pi^2 T_{SP})$. For
large fields,  a similar calculation can be done provided that
the field is smaller than the soliton gap and the  system does not
exhibit a transition to the incommensurate phase.

\begin{figure}[htbp]
  \begin{center}
    \includegraphics[width=\figwidth]{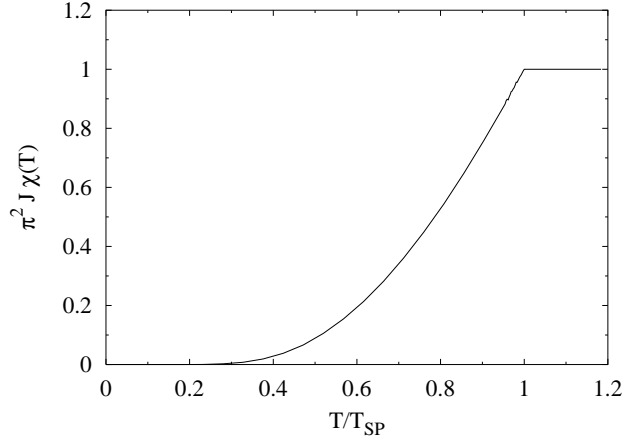}
    \caption{The magnetic susceptibility $\chi(T)$ versus the 
reduced temperature; for $T>T_{SP}$, $\chi(T)=1/(\pi^2 J)$. }
    \label{fig:magsuscep}
  \end{center}
\end{figure}

On the other hand to  calculate the finite temperature susceptibility, we need to generalize the TBA
equations
to include the effect of a magnetic field. To recapitulate,  in the
absence of a field, we have two solitons of spin $\pm
1$, a light breather of spin $0$ forming a triplet 
 and a heavy breather of spin $0$. A field breaks the spin degeneracy
and
for small enough fields which do not induce any incommensuration
we can use the TBA to calculate the magnetic susceptibility.
Following
Ref.~\onlinecite{konik_spinfield}, we obtain the following  TBA equations:
\begin{widetext}
\begin{eqnarray}
  \label{eq:tba-magfield}
  {\bar \epsilon}_{\sigma}(\theta)&=& \frac{\pi}{2} M \cosh \theta - {\bar h}\sigma +
  \frac {\bar T} {2\pi} \sum_{\sigma=-1,0,1} \int d\theta' 
  K_{11}(\theta-\theta')  \ln (1+e^{-{\bar \epsilon}_\sigma(\theta')/ {\bar T}}) 
 + \frac {\bar T} {2\pi}\int d\theta' 
  K_{12}(\theta-\theta')  \ln (1+e^{-{\bar \epsilon}_2(\theta')/{\bar T}}) \nonumber \\
{\bar \epsilon}_{2}(\theta)&=& \frac {\pi \sqrt{3}}{2} M \cosh \theta + \frac {\bar T} {2\pi} \sum_{\sigma=-1,0,1} \int d\theta'
  K_{12}(\theta-\theta')  \ln (1+e^{-{\bar \epsilon}_\sigma(\theta')/{\bar T}})
  +\frac {\bar T} {2\pi}\int d\theta' 
  K_{22}(\theta-\theta')  \ln (1+e^{-{\bar \epsilon}_2(\theta')/{\bar T}}) \nonumber  \\
f&=&\frac {\bar K} {2J} \delta^2 -\frac {\bar T} {2\pi} \int d\theta M \cosh
\theta \left[
\sum_{\sigma=1,0,-1} \ln (1+e^{-{\bar \epsilon}_\sigma(\theta)/{\bar T}}) +\sqrt{3}  \ln
(1+e^{-{\bar \epsilon}_2(\theta)/{\bar T}})\right] + \frac{M^2}{8 \sqrt{3}}  
\end{eqnarray}
\end{widetext}
\noindent where the ${ \bar \epsilon}_\sigma(\theta)$ denote the
reduced pseudoenergies of the
solitons ($\sigma=1$), antisolitons ($\sigma=-1$) and light breathers
($\sigma=0$) and $\bar{h}=h/J$. 
It is easy to see from the equations above that
${\bar \epsilon}_\sigma(\theta)={\bar \epsilon}_0(\theta)-{\bar h}\sigma$. This allows us to
reduce the set of TBA equations to  two:
\begin{widetext}
\begin{eqnarray}
  \label{eq:tba-mf-reduced}
 {\bar \epsilon}_0(\theta) &=& \frac{\pi}{2} M \cosh \theta +   \frac {3{\bar T}} {2\pi}  \int d\theta' 
  K_{11}(\theta-\theta')  \ln (1+e^{-{\bar \epsilon}_0(\theta')/{\bar T}}) + \frac {\bar T} {2\pi}\int d\theta' 
  K_{12}(\theta-\theta')  \ln (1+e^{-{\bar \epsilon}_2(\theta')/{\bar T}}) 
  \nonumber \\ && + \frac {\bar T}{2\pi} \int d\theta' 
  K_{11}(\theta-\theta')  \ln \left[1+\frac{\sinh^2 \left(\frac {\bar h}
{2{\bar T}}\right)}{\cosh^2\left(\frac{{\bar \epsilon}_0(\theta')}{2{\bar T}}\right)}\right] 
\nonumber   \\ 
{\bar \epsilon}_{2}(\theta)&=&\frac{\pi\sqrt{3}}{2} M  \cosh \theta + \frac {3{\bar T}} {2\pi}  \int d\theta' 
  K_{12}(\theta-\theta')  \ln (1+e^{-{\bar \epsilon}_0(\theta')/{\bar T}}) + \frac {\bar T} {2\pi}\int d\theta' 
  K_{22}(\theta-\theta')  \ln (1+e^{-{\bar \epsilon}_2(\theta')/{\bar T}})  \nonumber \\ && + \frac {\bar T}{2\pi} \int d\theta' 
  K_{12}(\theta-\theta')  \ln \left[1+\frac{\sinh^2\left(\frac {\bar h}
        {2{\bar T}}\right)}{\cosh^2\left(\frac{{\bar \epsilon}_0(\theta')}{2{\bar T}}\right)}\right]
\end{eqnarray}
\end{widetext}
In the presence of a magnetic field, the law of corresponding
states~(\ref{eq:corresp-states}) now
reads:
\begin{eqnarray}
  \label{eq:corresp-MF}
  F=-\frac{T_{SP}^2}{J}{\cal F}\left(\frac T {T_{SP}},\frac h {T_{SP}}\right)
\end{eqnarray}
The magnetic susceptibility,
$\chi(T)=-{\rm lim}_{h \to 0}\frac{\partial^2 F}{\partial h^2}$, satisfies the scaling relation 
\begin{eqnarray}
  \label{eq:corresp-chi}
  \chi(T)=\frac 1 J {\cal F}^{\prime \prime}_y\left(\frac T {T_{SP}},0\right),
\end{eqnarray}
\noindent where ${\cal F}^{\prime \prime}_y=\partial^2_y{\cal
  F}(x,y)$. 
For $T>T_{SP}$, the susceptibility is that of a free Bose gas:
$\chi(T)=\frac 1 {\pi^2 J}$.
For temperatures $0 \leq T \leq T_{SP}$, the susceptibility can be
obtained  numerically from (\ref{eq:tba-mf-reduced}) and
(\ref{eq:corresp-chi}). The results are plotted in Fig.~
\ref{fig:magsuscep}.

As in the case of zero magnetic field,  one has an effective  field
dependent Landau
Ginzburg
functional which describes the physics in the vicinity of the transition.
For $T \alt T_{SP}$ the behavior of the magnetic susceptibility is obtained
from the  following Landau-Ginzburg expansion: 
\begin{eqnarray}
  \label{eq:LG-magfield}
  F(T,h,\delta)=\frac {p} 2 (T-T_{SP}(h)) \delta^2 + \frac {q} 4 \delta^4
  -\frac{\chi_0}{2} h^2, 
\end{eqnarray}
\noindent where $\chi_0=\frac 1 {\pi^2 J}$ 
is the susceptibility of the undistorted
chain.
Minimizing $F$ with respect to $\delta$, one
finds for $T<T_{SP}$: 
\begin{eqnarray}
  \label{eq:magfield-min}
  F(T,h)=-\frac{p^2}{4q}(T-T_{SP}(0)+\lambda h^2)^2 -\frac{\chi_0}{2} h^2. 
\end{eqnarray}
The definition of $\chi$ then gives 
\begin{equation}\label{eq:suscep-LG}
\chi(T<T_{SP})=\chi_0 + \frac{p^2 \lambda}{q}(T-T_{SP}(0)).
\end{equation}
This behavior of the susceptibility is reminiscent of  the one seen 
in a N\'eel antiferromagnet\cite{landau_electrodynamics} which stems
from
the similarity of the 
 mean field
equations for the spin Peierls problem and the quasi-1D antiferromagnet. 
Using Eq. (\ref{eq:lambda-Tsp}) and
Eqs.~(\ref{eq:coeff-quadra}) it is easily seen that the resulting
susceptibility (\ref{eq:suscep-LG}) satisfies to the law of
corresponding states (\ref{eq:corresp-chi}). Numerically, one finds that
$\frac{p^2 \lambda}{q}=\frac{0.257}{J T_{SP}}$ by fitting the
susceptibility calculated near the transition to the Landau-Ginzburg
form. If on the other hand, we use the values of $p$ and $q$ obtained
 in Sec.~\ref{sec:LG-expans}  combined with the value of
$\lambda$ in Eq.~(\ref{eq:lambda-Tsp}) we see that value
$\frac{p^2\lambda}{q}=\frac{2.8\times 14.7}{16\pi^2 J
  T_{SP}}=\frac{0.26}{J T_{SP}}$  differs from that obtained in the
presence of a field by less than a percent. A Keesom-Ehrenfest
relation\cite{landau_statmech} 
exists between the jump of the specific heat and the jump in the slope
of the magnetic susceptibility as a function of temperature. 
This Keesom-Ehrenfest relation is given by the
Landau Ginzburg theory as: 
\begin{eqnarray}
  \label{eq:chi-Cv-LG}
  T_{SP}^2   \frac {\frac{d\chi}{dT}}  {\Delta C_v} =
  \frac{\pi^2-\psi^{(1)}(3/4)}{4\pi^2}=\frac{2\beta(2)}{\pi^2}\simeq 0.3724
\end{eqnarray}
Such a proportionality has been observed in experiments on
TTFAuBDT in magnetic fields.\cite{northby_ttfaubdt} 

\subsection{Low temperature expansions}
In the preceding sections, we have obtained analytical results for $T=0$ and
for $T\alt T_{SP}$. In fact, the TBA equations are amenable to
analytical study   
 for $T\agt 0$ (more precisely $0<T \ll \Delta(T=0)$). In this regime, we
expect that the mean field gap $\Delta(T)$ remains very close to
$\Delta(T=0)$, so that the thermodynamics does not differ from the one
of the Heisenberg chain with dimerization. 
With this assumption,  the TBA equations (for $h=0$) to lowest order are:
\begin{eqnarray}
  \label{eq:low-T-pseudo}
  \epsilon_1(\theta) & =&\Delta(T=0) \cosh \theta +O(e^{-\frac
    {\Delta(T=0)}T}) \nonumber  \\
\epsilon_2(\theta)& =& \Delta(T=0) \sqrt{3} \cosh \theta +O(e^{-\frac
    {\Delta(T=0)} T})
\end{eqnarray}
Substituting these in (\ref{eq:TBA-equations}) for  the free energy $F$, we see that the correction to $F$ is indeed $O(e^{-\Delta(T=0)/T})$ which justifies our original
assumption that $\Delta(T) \simeq \Delta(0)$. 
We now derive low temperature expansions of the various physical
quantities. It is convenient to use the
the zero temperature dimerization $\delta_0=\delta(T=0)$, the zero
temperature gap $\Delta_0=\Delta(T=0)$, and the zero temperature
groundstate energy $E_0$, the expressions of which are given in
(\ref{eq:zeroT}),  to express the corresponding finite temperature
quantities as a function of  the dimensionless variable
$\bar{\delta}$. We obviously have $\delta(T)=\delta_0
\bar{\delta}(T)$, and $\Delta(T)=\Delta_0 \bar{\delta}^{2/3}(T)$. The total energy at
$T=0$  reads
\begin{eqnarray}
  \label{eq:energy-deltabar}
  E(\bar{\delta})=\frac{4}{\sqrt{3}}
  \left(\frac{\Gamma(1/6)}{\Gamma(2/3)}\right)^6
  \left[\frac{\Gamma(3/4)}{\Gamma(1/4) \pi^2} \left(\frac
      \pi 2 \right)^{1/4} \right]^4 \frac{J^3}{k^2}
  \left[\bar{\delta}^2 -\frac {3}{2}\bar{\delta}^{\frac43}\right]
\end{eqnarray}
and it is easy to see that $ E(\bar{\delta})$ has a
minimum for $\bar{\delta}=1$. Expanding around this minimum we find:
\begin{eqnarray}
  \label{eq:energ-exp-delta}
  E(\bar{\delta})-E_0 &=&\frac{2}{9\pi\sqrt{3}} \frac{\Delta_0^2}{J}(\bar{\delta}-1)^2.
\end{eqnarray}
The expression (\ref{eq:energ-exp-delta}) is not the full expression
of the free energy for $T>0$, as we have also to take into account the
contributions of the solitons and the breathers that are thermally
excited. Since the heavy breathers have  mass $M\sqrt{3}$, as can
be seen from (\ref{eq:low-T-pseudo}), their contribution at low
temperature is negligible with respect to the soliton
contribution. Therefore, to lowest order, the thermal contribution to
the free energy reads:
\begin{eqnarray}
  \label{eq:thermal-lowT}
  F_{sol.}=-\frac{6T}{\pi^2} \frac{\Delta_0}{J} \bar{\delta}^{2/3}
  K_1\left(\frac{\Delta_0}{T} \bar{\delta}^{2/3}\right), 
\end{eqnarray}
\noindent where $K_1$ is a modified Bessel
function\cite{abramowitz_math_functions} 
so that the full variational free energy is:
\begin{eqnarray}
  \label{eq:fullF-lowT}
  F(T,\bar{\delta})=E_0 + \frac{2}{9\pi\sqrt{3}}
  \frac{\Delta_0^2}{J}(\bar{\delta}-1)^2 -\frac{6T}{\pi^2} \frac{\Delta_0}{J} \bar{\delta}^{2/3}
  K_1\left(\frac{\Delta_0}{T} \bar{\delta}^{2/3}\right). 
\end{eqnarray}
Minimizing (\ref{eq:fullF-lowT}) with respect to $\bar{\delta}$ we
obtain: 
\begin{eqnarray}
  \label{eq:deltabar-lowT}
  \bar{\delta}-1=\frac{9\sqrt{3}}{\pi} 
  \left[K_1'\left(\frac{\Delta_0}{T}\bar{\delta}^{2/3}\right) \bar{\delta}^{1/3}
      +\frac{T\bar{\delta}^{-1/3}}{\Delta_0}
      K_1\left(\frac{\Delta_0}{T}\bar{\delta}^{2/3}\right)\right] 
\end{eqnarray}
To lowest order, this gives:
\begin{eqnarray}
  \label{eq:deltabar-exp-lowT}
  \bar{\delta}=1-3^{7/2}\sqrt{\frac{2T}{\pi \Delta_0}} e^{-\frac
    {\Delta_0}{T}}, 
\end{eqnarray}
and:
\begin{eqnarray}
  \label{eq:gap-exp-lowT}
  \Delta(T)=\Delta_0\left[1- 6\sqrt{3} \sqrt{\frac{2T}{\pi \Delta_0}} e^{-\frac
    {\Delta_0}{T}}\right]. 
\end{eqnarray}
Substituting (\ref{eq:deltabar-exp-lowT}) in (\ref{eq:fullF-lowT}), we
see that the correction to the elastic energy plus ground state energy
of the dimerized chain is of order $e^{-2\Delta_0/T}$ is is
  therefore negligible compared to the contribution of the solitons. 
In physical terms, this means that at sufficiently low temperature,
the thermodynamics of the spin-Peierls chain is the same as the
thermodynamics of a chain with a constant dimerization. 
Using this result, we find that:
\begin{eqnarray}
  \label{eq:Free-lowT}
  F(T)=-\frac{\Delta_0^2}{J} \left[\frac{6 T}{\pi^2 \Delta_0}
    K_1\left(\frac{\Delta_0}{T}\right)\right], 
\end{eqnarray}
which leads to a low temperature specific heat of the form:
\begin{eqnarray}
  \label{eq:Cv-lowT}
  C_v(T) =\frac{3 \sqrt{2}}{\pi^{3/2}} \frac{\Delta_0}{J}
  \left(\frac{\Delta_0}{T}\right)^{3/2} e^{-\frac{\Delta_0}{T}}
  +o(T^{-3/2} e^{-\Delta_0/T}) 
\end{eqnarray}
In the presence of an infinitesimal applied magnetic field $h\ll T$, 
the lowest order contribution to the low temperature free energy is: 
\begin{eqnarray}
  \label{eq:Free-lowT-h}
  F(T)=-\frac{\Delta_0^2}{J} \left[\frac{2 T}{\pi^2 \Delta_0}
    K_1\left(\frac{\Delta_0}{T}\right)(1+2 \cosh \frac h T)\right], 
\end{eqnarray}
and the magnetic susceptibility is readily obtained as:
\begin{eqnarray}
  \label{eq:chi-lowT}
  \chi(T)=\frac{1}{\pi^2 J} \sqrt{\frac{8\pi \Delta_0}{T}}
  e^{-\frac{\Delta_0}{T}} + o(T^{-1/2} e^{-\Delta_0/T}) 
\end{eqnarray}

\section{Conclusions}
In this paper, we have studied the thermodynamics of the
spin-Peierls system treated within a mean-field approximation.
Using a combination of bosonization methods and the
thermodynamic Bethe ansatz, we have been able to obtain
quantitative results for the spin-Peierls transition temperature
$T_{SP}$, the spin-Peierls gap to triplet excitations, the
specific heat and magnetic susceptibility at arbitrary temperatures.
Our calculations are a quantitative improvement of the
results obtained by Cross and Fisher (who were restricted to the
vicinity of the spin-Peierls transition temperature) and
consequently help us obtain the effective Landau Ginzburg
functional that describes the physics of dimerization close
to the transition. 
It would be interesting to study this Landau Ginzburg theory 
in one dimension, following Ref.~\onlinecite{bourbonnais96_spinpeierls} 
to understand more quantitatively 
how lattice fluctuations affect the thermodynamics. Similarly to
Ref.\onlinecite{bourbonnais96_spinpeierls}, we should expect a regime of
renormalized Gaussian fluctuations for $0.4T_{SP}^{MF}<T<T_{SP}^{MF}$,
and a regime dominated by kinks for $0.3T_{SP}<T<0.4T_{SP}$ as long as
one dimensional fluctuations dominate. Also, it should be possible to
use the Landau-Ginzburg description to study the three-dimensional
ordering of the dimerization along the lines of
Refs.~\onlinecite{scalapino_1d_2d,scalapino_q1d}. The dependence of
the transition temperature on interchain coupling will be similar to
the one predicted in Ref.~\onlinecite{mostovoy97_soliton_peierls}
since both models belong to the same universality class.   
A more direct extension of the present work would  be to study the
commensurate-incommensurate transition driven by an external
magnetic field and then comparing the predicted results to
various experiments on spin-Peierls systems. It would also be
interesting to extend this study to the generalized spin-Peierls
transition obtained in ladders under magnetic
field\cite{nagaosa_lattice_ladder} or to the antiferromagnetic phase
transition obtained in the same
system.\cite{giamarchi_coupled_ladders} These questions are left for
future work.

\begin{acknowledgments}
We thank R. Citro, T. Giamarchi, P. Lecheminant, F. H. L. Essler and
N. Andrei  for discussions.
\end{acknowledgments}
\appendix
\section{Calculation of the rigidity}\label{sec:calculation-rigidity}
Here, we present a derivation  of the rigidity in
our Landau-Ginzburg effective action. 
In the continuum limit, the  space dependent dimerization
leads to the modification of the
sine Gordon term in (\ref{eq:ham})
\begin{equation}
H_{int}= -\frac{2g}{(2\pi
    a)^2} \int dx \delta(x) \cos \sqrt{2}\phi.
\end{equation}
Close to the transition, the  
second order correction to the  free energy of the spin
chain induced by the spin-phonon coupling is given by:
\begin{equation} \label{eq-grad}
F_\delta = \frac 1 4 \frac{\pi
    a}{\beta u} \left(\frac{g}{\pi
      a}\right)^2 \int_{-\infty}^\infty dx \int_{-\infty}^\infty
    dx^\prime \int_0^\beta
\delta(x)\delta(x^\prime) \chi(x - x^\prime, \tau),
\end{equation}
and:
 \begin{equation}
\chi(x - x^\prime, \tau)=  \sqrt{2}\left[\cosh \frac{2\pi x}{\beta u}-\cos \frac {2\pi
        \tau}{\beta}\right]^{-\frac 12}.
\end{equation}
In Fourier space, 
$F_\delta \propto  \int \frac{dq}{2\pi} \delta(q)\delta(-q) \hat{\chi}(q,i\omega_n=0)$.
To obtain the gradient term, it thus suffices to calculate the Fourier
transform $\hat{\chi}$. In the limit  $q \to 0$, $\hat{
  \chi}(q,i\omega_n=0)= \hat{\chi}(0,0) + q^2/2 \hat{\chi}^{\prime
  \prime} (0,0)$. Plugging this form into (\ref{eq-grad}), it is
straightforward to find the rigidity. 

To find $\hat{\chi}(q)$ we
generalize slightly the calculation of the $q=0$ response function, and
consider:
\begin{eqnarray}
  \label{eq:suscep-q-dependent}
 \hat{\chi}(q)&=& \int dx d\tau \chi(x,\tau) e^{iqx}  \nonumber \\
&=& \int dx d\tau
  \frac{\sqrt{2}}{\sqrt{\left[ \cosh \left(\frac{2\pi
          x}{\beta u}\right) -\cos \left(\frac{2\pi
          \tau}{\beta}\right)\right]}} e^{iqx}.
\end{eqnarray}
\noindent Using Eq. (8.12.5) in
Ref.~\onlinecite{abramowitz_math_functions}, we can rewrite the 
integral:
\begin{eqnarray}
  \frac a 2 \int_{-\infty}^\infty dv \frac{e^{\frac{iq\beta u}{2\pi}
      v}}{\sqrt{\cosh v -\cos \frac{2\pi \tau}{\beta}}} =  
  \frac{\pi}{\sqrt{2}} a \frac 1 {\cosh \left(\frac{q \beta
        u}{2}\right)} P_{-1/2+ i\frac{\beta uq}{2\pi}}\left(-\cos
    \frac {2\pi\tau}{\beta}\right). 
\end{eqnarray}
Consequently, to calculate (\ref{eq:suscep-q-dependent}), we only need
the integral:
\begin{eqnarray}
  \int_0^\beta P_{-1/2 + i\frac{\beta uq}{2\pi}}\left(-\cos \frac{2\pi
      \tau}{\beta}\right) d\tau =\frac{\pi \beta}{[\Gamma\left(\frac 3
      4 + \frac {iq u}{4\pi T}\right)  \Gamma\left(\frac 3
      4 - \frac {iq u}{4\pi T}\right)]^2},
\end{eqnarray}
\noindent which is easily obtained from Eq. (8.14.16) in
Ref.~\onlinecite{abramowitz_math_functions}. The final result is:
\begin{eqnarray}
  \hat{\chi}(q)=\frac{\pi^2}{2\cosh\left(\frac {u q}{2T}\right) \Gamma\left(\frac 3
      4 + \frac {iq u}{4\pi T}\right)^2  \Gamma\left(\frac 3
      4 - \frac {iq u}{4\pi T}\right)^2}.
\end{eqnarray}
\noindent
Expanding $\hat{\chi}(q)$ to second order in $q$, we find:
\begin{eqnarray}
  \label{eq:LG-gradient}
  F=F_0-\frac{1}{16\pi^2 a^3 T \Gamma(3/4)^4} \int
  \frac{dq}{2\pi} |g(q)|^2
  \left[1-\eta
    \left(\frac{uq}{2T}\right)^2\right], 
\end{eqnarray}
\noindent where 
\begin{eqnarray}
  g(q)=\int g\delta(x) e^{iqx} dx,
\end{eqnarray}
\noindent and $\eta=\frac{\pi^2-\psi^{(1)}(3/4)}{2\pi^2}$ and  $\psi^{(1)}(x)$ is the trigamma function (see
Ref.~\onlinecite{abramowitz_math_functions} p.260). The number $\psi^{(1)}(3/4)$ can be
expressed as a function of Catalan's
constant\cite{abramowitz_math_functions} $\beta(2)$ as 
$\psi^{(1)}(3/4)=\pi^2-8\beta(2)\simeq 2.5419$. 
Finally, using the expression of $g(x)$ as a function of $\delta(x)$
we obtain the rigidity $c_0$ as:
\begin{eqnarray}
  \label{eq:rigidity}
  c_0& =&\frac{9}{64}\left(1-\frac{\psi^{(1)}(3/4)}{\pi^2}\right)\left(\frac
    \pi 2\right)^{1/2} \frac 1 {\Gamma(3/4)^4}
  \frac{J^4 a}{T^3} \nonumber \\
& =&\frac{9}{8\pi^2} \left(\frac \pi 2\right)^{1/2}
  \frac {\beta(2)} {\Gamma(3/4)^4} \frac{J^4 a}{T^3}
\end{eqnarray}

A similar calculation can be done to obtain  the reduction of the critical
temperature as a function of the magnetic field. When the system is
magnetized, incommensuration in the staggered operator sets in  as the
wavevector shifts from $\pi/2$ to
$\pi/2\pm h/u$
and  the equation giving the critical temperature reads:
\begin{eqnarray}
  \label{eq:Tc-magfield}
  k=\frac{a}{2} \left(\frac{g}{2 a\pi}\right)^{2} \frac
  {\beta  \sech \left( \frac {\beta h} 2\right)}{ \Gamma\left(\frac 3
      4 -i  \frac{\beta h}{4\pi}\right)^2     \Gamma\left(\frac 3 4 +i  \frac{\beta h}{4\pi}\right)^2},    
\end{eqnarray}
\noindent
This implies that:
\begin{widetext}
\begin{eqnarray}
  \label{eq:Tc-relative}
  T_{SP}(h)\cosh \left(\frac h
      {2T_{SP}(h)}\right)\left[  \Gamma\left(\frac 3
      4 -i  \frac{h}{4\pi T_{SP}(h)}\right) \Gamma\left(\frac 3
      4 +i  \frac{h}{4\pi T_{SP}(h)}\right)\right]^2 = T_{SP}(h=0) \Gamma(3/4)^4.
\end{eqnarray}
\end{widetext}
\noindent the Equation (\ref{eq:Tc-relative}) was obtained in
Ref.~\onlinecite{cross_field_sP} using a real time calculation of the response
function. This can be seen explicitly by expressing the infinite
products in Ref.~\onlinecite{cross_field_sP} in terms of Gamma functions. 
Expanding  (\ref{eq:Tc-relative}) around small $h$, one obtains for magnetic
fields $h\ll T_{SP}(0)$:
\begin{eqnarray}
  \label{eq:field-correction-Tsp}
  \frac{T_{SP}(h)}{T_{SP}(0)}\simeq 1-2(\pi^2 -\psi^{(1)}(3/4)) \left(\frac
    h{4\pi T_{SP}}\right)^2.  
\end{eqnarray}
In terms of  Catalan's constant, the spin-Peierls transition
temperature in the presence of a small field is given by: 
\begin{eqnarray}
  \frac{T_{SP}(h)}{T_{SP}(0)}=1-\beta(2) \left(\frac
    h{\pi T_{SP}}\right)^2+o(h^2).
\end{eqnarray}

%\bibliographystyle{phaip}
%\bibliography{revues,1d,general,spinpeierls,nrefs,math,perso,books}

\end{document}